\documentclass[twocolumn,showpacs,aps,groupedaddress,prd,eqsecnum]{revtex4}
\usepackage{amsfonts,latexsym,eucal}
\usepackage[dvips]{graphicx}

\def\beq{\begin{equation}}
\def\eeq{\end{equation}}
\def\bea{\begin{eqnarray}}
\def\eea{\end{eqnarray}}
\def\nn{\nonumber\\}
\def\pa{\partial}
\def\ra{\rightarrow}
\def\mn{_{\mu\nu}}

\def\tt{\tilde{t}}
\def\tr{\tilde{r}}
\def\tp{\tilde{\phi}}
\def\tpp{{\tp_{~}}'}
\def\tpps{{\tp_{~}}^{\prime2}}
\def\tv{\tilde{V}}

\def\tm{\tilde{m}}
\def\tmu{\tilde{\mu}}
\def\to{\tilde{\omega}}
\def\bp{\mbox{\boldmath$\phi$}}

\def\tE{\tilde{E}}
\def\tQ{\tilde{Q}}
\newcommand{\dalm}{\kern1pt\vbox{\hrule height 0.9pt\hbox{\vrule width 0.9pt
\hskip 2.5pt\vbox{\vskip 5.5pt}\hskip 3pt\vrule width 0.3pt}\hrule height 0.3pt}\kern1pt}

\usepackage{epsfig}

\begin{document}

\thispagestyle{empty}

\title{Unified picture of Q-balls and boson stars via catastrophe theory}
\author{Takashi Tamaki}
\email{tamaki@ge.ce.nihon-u.ac.jp}
\affiliation{Department of Physics, General Education, College of Engineering, 
Nihon University, Tokusada, Tamura, Koriyama, Fukushima 963-8642, Japan}
\author{Nobuyuki Sakai}
\email{nsakai@e.yamagata-u.ac.jp}
\affiliation{Department of Education, Yamagata University, Yamagata 990-8560, Japan}

\begin{abstract}
We make an analysis of Q-balls and boson stars using catastrophe theory,
as an extension of the previous work on Q-balls in flat spacetime.
We adopt the potential $V_3(\phi)={m^2\over2}\phi^2-\mu\phi^3+\lambda\phi^4$ for Q-balls and 
that with $\mu =0$ for boson stars.
For solutions with $|g^{rr}-1|\sim 1$ at its peak, stability of Q-balls has 
been lost regardless of the potential parameters. As a result, phase relations, such as 
a Q-ball charge versus a total Hamiltonian energy, approach those of boson stars, 
which tell us an unified picture of Q-balls and boson stars. 
\end{abstract}

\maketitle

\section{Introduction}

Among non-topological solitons which appear in $U(1)$-symmetric scalar fields, objects existing even in flat spacetime are called Q-balls \cite{LP92,Col85,SUSY,SUSY-DM,Kus97b-98,Kus97a,PCS01, Kusmartsev,SakaiSasaki,Copeland,Volkov}, while objects supported by strong gravity are called boson stars \cite{boson-review,axidilaton,Mielke,memory,KMS91,KKL05}.

Q-balls are typically supposed to be microscopic objects. It has been argued that Q-balls generally exist 
in all supersymmetric extensions of the Standard Model and could play important roles in 
cosmology \cite{Kus97b-98}.  For example, Q-balls can be produced efficiently in the Affleck-Dine mechanism and could be responsible for baryon asymmetry \cite{SUSY} and dark matter \cite{SUSY-DM}.
As for the stability of Q-ball in flat spacetime, analytic results were obtained under the thin-wall approximation  \cite{Col85} and under the thick-wall approximation \cite{Kus97a}.
A general criterion of stability was derived analytically by Pacceti Correia and Schmidt \cite{PCS01}:
a Q-ball is stable if and only if 
\beq\label{Qw-condition}
\frac{\omega}{Q}\frac{dQ}{d\omega}<0\ ,
\eeq
where $\omega$ and $Q$ are the angular velocity of phase rotation and the Q-ball charge, respectively. 
In general cases, numerical calculation is necessary for having equilibrium solutions, and catastrophe theory \cite{PS78} is a useful tool for finding their stability. 
Kusmatsev made a general argument that catastrophe theory can be applied to the investigation of 
molecules and solitons in various systems and found the criterion (\ref{Qw-condition})~\cite{Kusmartsev}.
In  \cite{SakaiSasaki} catastrophe theory was applied to two typical models of Q-balls to explore the stability for the whole parameter space, which includes the intermediate region between the thin-wall limit and thick-wall limit.
In this paper, as well as in most of the previous works, stability means local stability, that is, stability against small perturbations. Absolute stability of Q-balls has been argued in \cite{Copeland}.

On the other hand, boson stars have been studied as astronomical objects which could also 
contribute to dark matter~\cite{boson-review}. For example, 
axidilaton star of $\sim 0.6M_{\odot}$ could account for a part of massive compact halo objects~\cite{axidilaton}.  
Supermassive boson stars of $10^{6}$-$10^{9}M_{\odot}$ have been discussed as an alternative to a black hole in the galaxy 
center~\cite{Mielke}. 
If we consider the evolution of boson stars in scalar-tensor theories, they could have gravitational memory: the 
strength of the gravitational constant at formation time could still be effective \cite{memory}. 
As for the stability analysis of boson stars, catastrophe theory has also been used \cite{KMS91}. 

We may note, in passing, that catastrophe theory has also been applied to the stability analysis of non-Abelian black holes \cite{Maeda}.

Although  the difference in theory between Q-balls and boson stars is solely the model parameters, the investigations of 
their properties have been carried out separately.  
This is because Q-balls and boson stars have been discussed in different contexts of particle physics or astrophysics.
The purpose of the present paper is to obtain an unified picture of equilibrium solutions and their stability of Q-balls and boson stars.
Our interest is essentially mathematical. 
However, gravitating Q-balls, or Q-stars, which are intermediate objects between Q-balls in flat spacetime and boson stars, have also been discussed \cite{FLP87,Grav-Q,multamaki}.
Therefore, an unified analysis of Q-balls and boson stars is also important for the study of astrophysical models.

This paper is organized as follows.
In Sec. II, we derive equilibrium field equations and explain how to analyze stability using catastrophe theory. 
In Sec. III, we show numerical results of equilibrium Q-balls and boson stars and discuss their stability.
In Sec. IV, we devote to concluding remarks.

\section{Analysis method of equilibrium Q-balls and boson stars}

\subsection{Equilibrium field equations}

We begin with the action
\bea\label{Sg}
{\cal S}&=&\int d^4x({\cal L}_{\rm G}+{\cal L}_{\phi}), 
\nonumber  \\
{\cal L}_{\rm G}&\equiv &\sqrt{-g}{{\cal R}\over 16\pi G}\ ,
\nonumber  \\
{\cal L}_{\phi}&\equiv &\sqrt{-g}\left\{-\frac12g^{\mu\nu}\pa_{\mu}\bp\cdot\pa_{\nu}\bp
-V(\phi) \right\},
\eea
where $\bp=(\phi_1,~\phi_2)$ is a SO(2)-symmetric scalar field and 
$\phi\equiv\sqrt{\bp\cdot\bp}=\sqrt{\phi_1^2+\phi_2^2}$.
We assume a spherically symmetric and static spacetime, 
\beq\label{metric1}
ds^2=-\alpha^2(r)dt^2+A^2(r)dr^2+r^2(d\theta^2+\sin^2\theta d\varphi^2).
\eeq

For the scalar field, we assume that it has a spherically symmetric and stationary form, 
\beq\label{phase}
(\phi_1,\phi_2)=\phi(r)(\cos\omega t,\sin\omega t).
\eeq
Then the field equations become
\bea\label{Gtt}
-{r A^3\over2}G^t_t&\equiv&A'+{A\over2r}(A^2-1) \nonumber \\
&=&{4\pi G}r A^3\left({{\phi'}^2\over2A^2}
+{\omega^2\phi^2\over2\alpha^2}+V\right),
\\\label{Grr}
{r\alpha\over2}G_{rr}&\equiv&\alpha'+{\alpha\over2r}(1-A^2) \nonumber \\
&=&{4\pi G}r\alpha A^2
\left({{\phi'}^2\over2A^2}+{\omega^2\phi^2\over2\alpha^2}-V\right),
\\\label{Box}
{A^2\phi\over\phi_1}\Box\phi_1&\equiv&
\phi''+\left(\frac2r+{\alpha'\over\alpha}-{A'\over A}\right)\phi'
+\left({\omega A\over\alpha}\right)^2\phi \nonumber \\
&=&A^2{dV\over d\phi},
\eea
where $'\equiv d/dr$. 
To obtain Q-ball solutions in curved spacetime, we should solve 
(\ref{Gtt})-(\ref{Box}) with boundary conditions, 
\bea
&& A(0)=A(\infty)=\alpha(\infty)=1,\nonumber \\ 
&& A'(0)=\alpha'(0)=\phi'(0)=\phi(\infty)=0.
\label{bcg}
\eea
We also restrict our solutions to monotonically decreasing $\phi (r)$. 
Due to the symmetry, there is a conserved charge called Q-ball charge,
\bea\label{Q}
Q&\equiv &\int d^3x\sqrt{-g}g^{\mu\nu}(\phi_1\pa_\nu\phi_2-\phi_2\pa_\nu\phi_1)=\omega I,
\nonumber  \\
&&{\rm where}~~~
I\equiv4\pi\int{A r^2\phi^2\over\alpha}dr.
\eea

\subsection{Models of Q-balls and boson stars}

As for Q-balls, which are present even in flat spacetime, we suppose the potential,
\beq\label{V3}
V_{3}(\phi):={m^2\over2}\phi^2-\mu\phi^3+\lambda\phi^4 ~~~
{\rm with} ~~~ m^2,~\mu,~\lambda>0\ ,
\eeq
which we call $V_3$ Model. Rescaling the quantities as~\cite{footnote}
\bea
\tt\equiv mt,~~ 
\tr\equiv mr,
&&\to\equiv{\omega\over m},~~
\tilde{\mu}\equiv{\mu\over\sqrt{\lambda}m},~~
\kappa\equiv{m^2G\over\lambda},\nn
\tilde{\phi}\equiv{\sqrt{\lambda}\over m}\phi,~~
&&\tv\equiv{\lambda\over m^4}V_{3}={\tilde{\phi}^2\over2}-\tilde{\mu}\tp^3+\tp^4,\nn
\tQ\equiv \lambda Q,
\label{rescale}
\eea
the field equations (\ref{Gtt}), (\ref{Grr}) and (\ref{Box}) with the potential 
(\ref{V3}) are rewritten as
\beq\label{rsfe1}
A'+{A\over2\tr}(A^2-1)
=4\pi\kappa\tr A^3\left({\tpps\over2A^2}+{\to^2\tp^2\over2\alpha^2}+\tv\right),
\eeq\beq\label{rsfe2}
\alpha'+{\alpha\over2\tr}(1-A^2)
=4\pi\kappa\tr\alpha A^2\left({\tpps\over2A^2}+{\to^2\tp^2\over2\alpha^2}-\tv\right),
\eeq\beq\label{rsfe3}
\tp^{\prime\prime}+\left(\frac2{\tr}+{\alpha'\over\alpha}-{A'\over A}\right)\tpp
+\left({\to A\over\alpha}\right)^2\tp=A^2{d\tv\over d\tp}.
\eeq

For reference, we recall the parameter regions of $\tilde{\omega}^{2}$ where solutions exist 
in flat spacetime. In this case, the field equation is 
\beq\label{fe}
\frac{d^2\tilde{\phi} }{ d\tilde{r}^2}=-\frac{2}{\tilde{r}}\frac{d\tilde{\phi}}{ d\tilde{r}}-
\tilde{\omega}^2\tilde{\phi}+{d\tilde{V}\over d\tilde{\phi}}\,.
\eeq
This is equivalent to the field equation for a single static scalar 
field with the potential $V_{\omega}\equiv \tilde{V}-\tilde{\omega}^2\tilde{\phi}^2/2$.
Equilibrium solutions satisfying boundary conditions (\ref{bcg}) 
exist if min$(V_{\omega})<\tilde{V}(0)$ and $d^2V_{\omega}/d\tilde{\phi}^2(0)>0$,
which is equivalent to
\beq\label{omega}
1-{\tilde{\mu}^2\over2}<\to^2<1.
\eeq
The two limits $\tilde{\omega}^2\ra1-{\tilde{\mu}^2\over2}$ and $\tilde{\omega}^2\ra 1$ 
correspond to the thin-wall limit and the thick-wall limit, respectively. 
As we shall explain below, qualitative features of solutions change at $\tilde{\mu}^{2}=2$. 

On the other hand, as for boson stars, we assume the potential (\ref{V3}) with $\mu =0$. 
Their equilibrium solutions do not exist in flat spacetime. 

The condition (\ref{omega}) cannot apply to gravitating solutions. Actually, boson stars exist.
However, as we shall show below, the criterion $\to^2<1$ is valid even for 
gravitating Q-balls and boson stars.

\subsection{Stability analysis method via catastrophe theory}

Let us discuss how we apply catastrophe theory to the present Q-ball or boson star system. 
An essential point is to choose {\it behavior variable}({\it s}),
{\it control parameter}({\it s}) and a {\it potential\/} in the Q-ball or boson star system appropriately. 

In \cite{SakaiSasaki} we argued that the total energy of the scalar field,
\beq
E_{\phi}\equiv\int d^3x\left\{{\omega^2\phi^2\over2}+{(\phi')^2\over2}+V\right\},
\eeq
is appropriate for a {\it potential\/} because  the variation of $E_{\phi}$ under fixed $Q$, 
$\delta E_{\phi}/\delta\phi|_Q = 0$, reproduces the equilibrium field equation (\ref{fe}). 
This is on the analogy of a {\it potential} in a mechanical system, where the potential $F(x)$ is defined in such a way that $dF/dx=0$ at equilibrium points.
 
A nontrivial issue in curved spacetime is the choice of the corresponding total energy since there are many definitions for total energy. However, we can conclude that the Hamiltonian energy $E$, which we shall calculate below, is the appropriate because $\delta E/\delta\phi|_Q = \delta E/\delta g\mn=0$, reproduces the equilibrium field equations (\ref{Gtt})-(\ref{Box}).

From the Lgrangian (\ref{Sg}) with the coordinate system (\ref{metric1}), one finds the 
canonical momentum of $\phi$ and the matter part of the Hamiltonian density,
\beq
{\cal P}_a={\pa{\cal L}_{\phi}\over\pa\dot{\phi}_a}={\sqrt{-g}\over\alpha^2}\dot\phi_a,~~~
\sqrt{-g}=\alpha Ar^2\sin\theta,
\eeq\beq
{\cal H}_{\phi}={\cal P}_a\dot\phi_a-{\cal L}_{\phi}
={\alpha^2{\cal P}_a^2\over2\sqrt{-g}}+\sqrt{-g}\left\{{(\phi_a')^2\over2A^2}+V\right\}.
\eeq
where $\dot{~}\equiv\pa/\pa t$.
Under the stationary condition (\ref{phase}), we obtain
\beq
{\cal H}_{\phi}=\sqrt{-g}\left\{{\omega^2\phi^2\over2\alpha^2}+{(\phi')^2\over2A^2}+V\right\}
\equiv\sqrt{-g}\rho_{\phi}.
\eeq
Similarly, we can consider the canonical momentum of the spatial metric $h_{ij}$ and the 
gravity part of the Hamiltonian density, 
\beq
{\cal \pi}^{ij}={\pa{\cal L}_{\rm G}\over\pa \dot{h}_{ij}},
\eeq\bea
{\cal H}_{\rm G}&=&{\cal \pi}^{ij}\dot h_{ij}-{\cal L}_{\rm G}
=-\sqrt{-g}{{\cal R}\over16\pi G}\nn
&=&{\sqrt{-g}\over8\pi G}\left\{G^t_t+{1\over r^2\alpha A}\left({r^2\alpha'\over A}\right)'\right\},
\eea
where we have used the static condition, $\dot h_{ij}=0$. 
Using one of the field equations, $G^t_t=-8\pi G\rho_{\phi}$, we obtain the total Hamiltonian,
\beq\label{H}
E\equiv\int d^3x({\cal H}_{\rm G}+{\cal H}_{\phi})
=\lim_{r\ra\infty}{r^2\alpha'\over2GA}.
\eeq
If we define the gravitational mass $M$ by the asymptotic behavior of the metric,
\beq
\alpha^2,~A^{-2}\ra1-{2GM\over r}
~~~ {\rm as} ~~~ r\ra\infty,
\eeq
Eq.(\ref{H}) reduces to
\beq\label{E}
E={M\over2}.
\eeq
In the previous work \cite{FLP87} the total Hamiltonian was calculated as $E=M$;
however, what we have shown is that the correct formula is (\ref{E}). 
We also use the normalized quantity 
\beq\label{Enormalize}
\tE\equiv{\lambda\over m}E.
\eeq

Because the charge $Q$ and the model parameter(s) of $V(\phi)$ can be given by hand, they should be 
regarded as {\it control parameters}. In flat spacetime, $V_3$ Model essentially has only one parameter, 
$\tilde{\mu}^2$. In curved spacetime, on the other hand, the normalized gravitational constant $\kappa$ is another {\it control parameter}, which represents the strength of gravity.

To discuss a {\it behavior variable} we consider an one-parameter family of perturbed field configurations $\phi_{x}(r)$ near the equilibrium solution $\phi(r)$.
Because $dE[\phi_x]/dx=(\delta E/\delta\phi_x)d\phi_x/dx= 0$ when $\phi_{x}$  is an equilibrium solution, $x$ is  a {\it behavior variable}. 
Although an explicit choice for $x$ is not unique, we choose $\to^2$ as a {\it behavior variable}. 

According to Thom's theorem, if the system has two control parameters in a mechanical system, 
there is essentially one behavior variable;  if the system has three control parameters, there are one or 
two behavior variables. Because the present Q-ball or boson star system 
contains $(\tilde{Q}, \tilde{\mu}^2, \kappa)$, we speculate that each has two {\it behavior variables}, 
$\tilde{\omega}^{2}$ and $\tilde{\phi}(0)$, and falls into {\it hyperbolic umbilical} catastrophe.
However, because the stability structure of equilibrium solutions in three-parameter space 
$(\tilde{Q}, \tilde{\mu}^2, \kappa)$ 
is very complicated and our interest is how gravitational effects change the stability structure, in the following, 
we discuss the stability structure of equilibrium solutions in two-parameter space $(\tilde{Q}, \kappa)$ 
under fixed $\tilde{\mu}^2$.

\begin{figure}[htbp]
\psfig{file=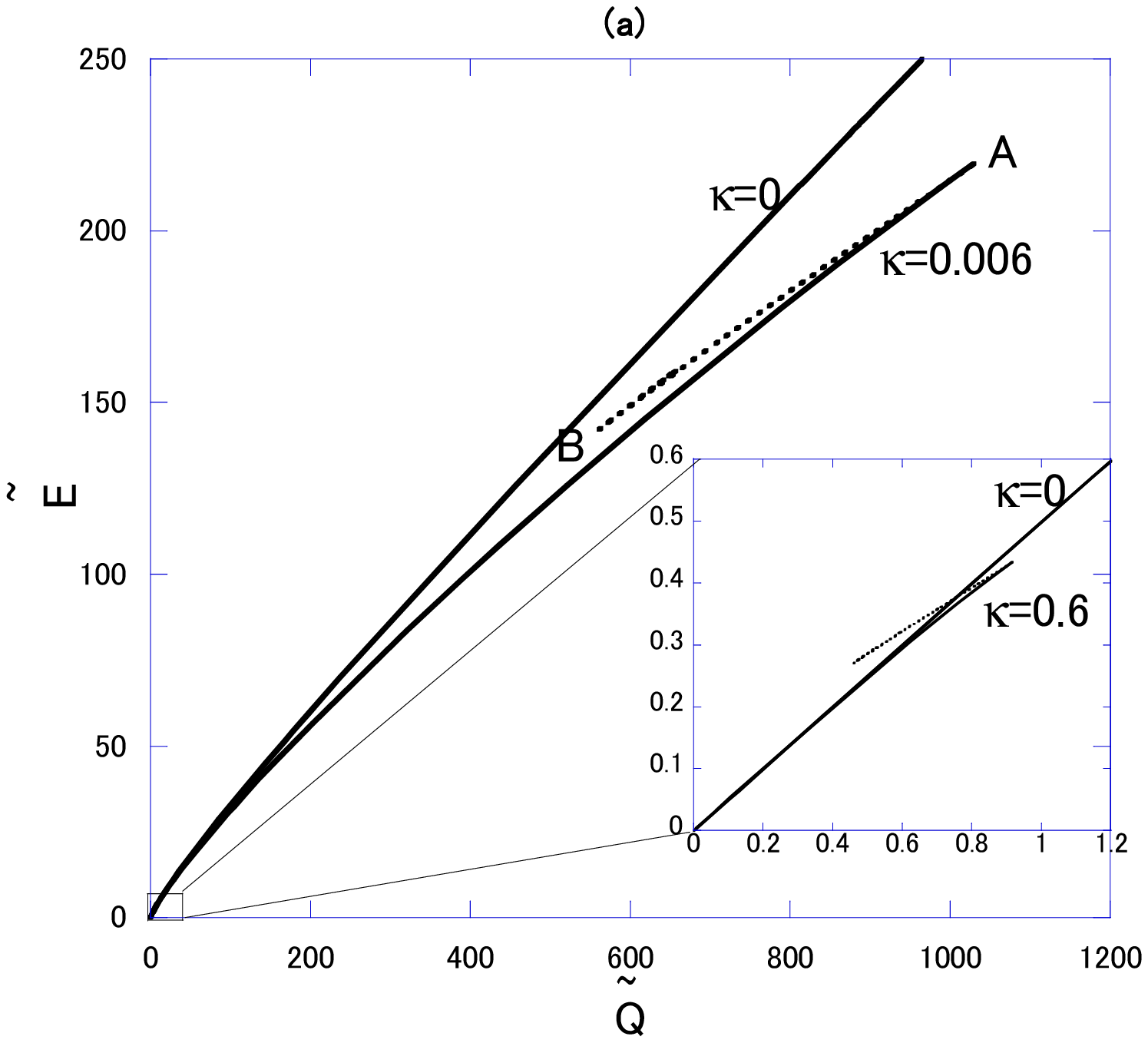,width=3in}  \\
\psfig{file=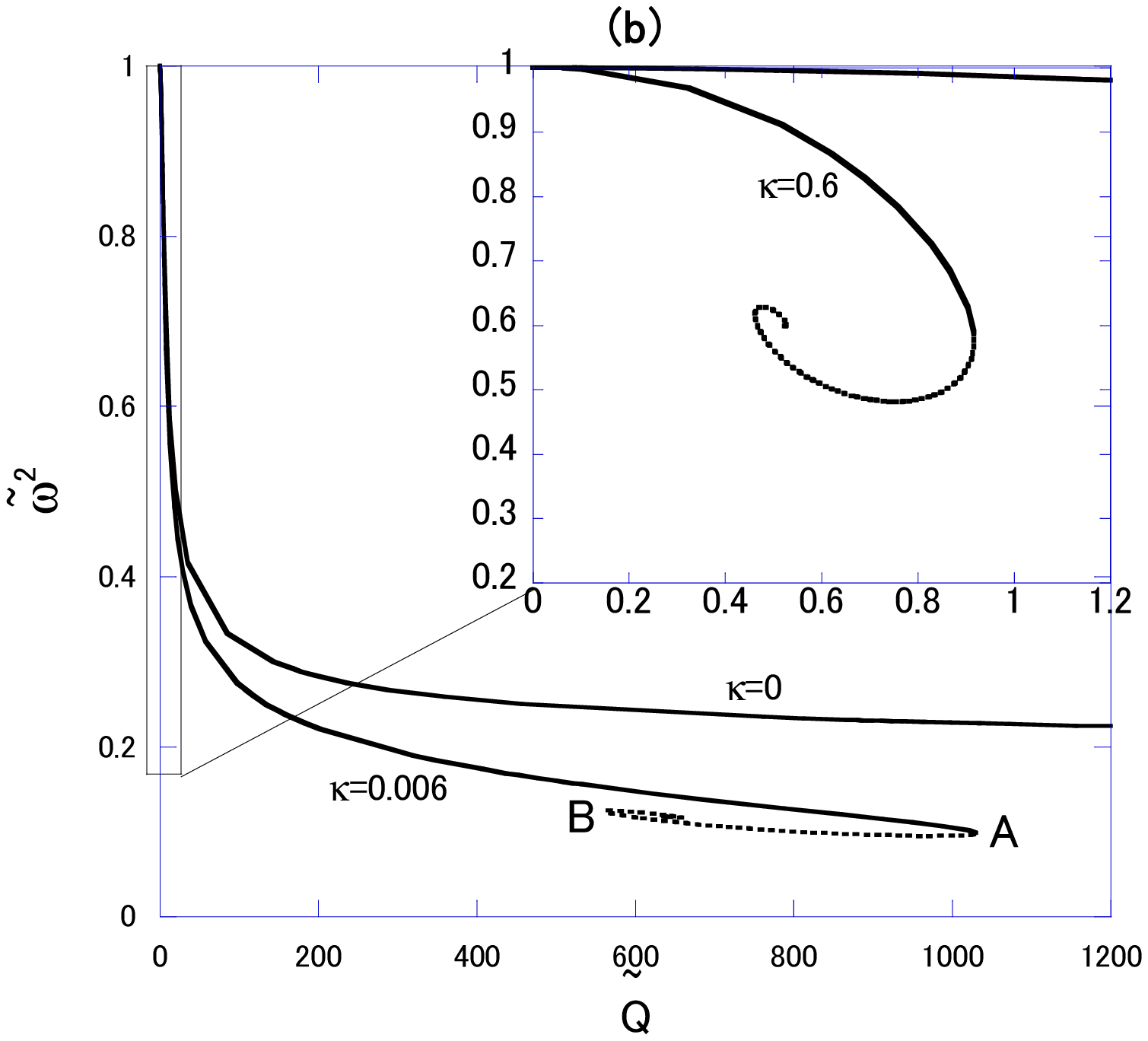,width=3in}
\caption{(a) $\tilde{Q}$-$\tilde{E}$ relation and (b) $\tilde{Q}$-$\tilde{\omega}^{2}$ relation for 
$\tilde{\mu}^{2}=\frac{5}{3}$ in $V_{3}$ Model. We compare solutions for $\kappa=0$, $0.006$ and $0.6$. 
In the case of gravitating Q-balls ($\kappa\ne0$) cusp and spiral structures can be seen in (a) and (b), 
respectively. 
The maximum of $\tQ$ and the local minimum value of $\tQ$ for $\kappa=0.006$ are labeled as A and B, respectively.
We interpret solutions with solid (dotted) lines stable (unstable). 
\label{QEw2ml05}}
\end{figure}

Our method of analyzing the stability of Q-balls and boson stars is as follows.
\begin{itemize}
\item Fix the value of $\tilde{\mu}^2$.
\item
Solve the field equations~(\ref{Gtt})-(\ref{Box}) with the boundary 
condition~(\ref{bcg}) numerically to obtain equilibrium 
solutions $\tilde{\phi}(r)$ for various values of $\tilde{\omega}$ and $\kappa$.
\item 
Calculate $\tilde{Q}$ for each solution to obtain the {\it equilibrium space} 
${\cal M}=\{(x,\tilde{Q},\kappa)\}$.
We denote the equation that determines ${\cal M}$ by $f(x,\tilde{Q},\kappa)=0$.
\item 
Find folding points where $\pa \tilde{Q}/\pa x=0$ or $\pa\kappa/\pa x=0$, 
in ${\cal M}$, which are identical to the stability-change points, 
$\Sigma=\{(x,\tilde{Q},\kappa)\,|\,{\pa f/\pa x}=0, ~f=0\}$.
\item
Calculate the energy $\tilde{E}$ by (\ref{H}) for equilibrium solutions
around a certain point in $\Sigma$ to find whether the point is 
a local maximum or a local minimum. Then we find the stability 
structure for the whole ${\cal M}$.
\end{itemize}

\section{Stability of gravitating Q-balls and boson stars}
Let us consider Q-balls of $V_3$ Model (\ref{V3}).
In flat spacetime ($\kappa=0$), stability structure falls into two classes,  
$\tilde{\mu}^2 < 2$ and $\tilde{\mu}^2 >2$ \cite{SakaiSasaki}:
\begin{itemize}
\item $\tilde{\mu}^2< 2$: All equilibrium solutions are stable. 
\item $\tilde{\mu}^2>2$ : For each $\tilde{\mu}^2$, there is a maximum charge, 
$\tQ_{\rm max}$, above which equilibrium solutions do not exist. 
For $\tQ<\tQ_{\rm max}$, stable and unstable solutions coexist. 
\end{itemize}

\subsection{Gravitating Q-balls for $\tilde{\mu}^2 < 2$}

In this subsection, we fix $\tilde{\mu}^2 =\frac{5}{3}$ as an example of $\tilde{\mu}^2 < 2$.
Figure \ref{QEw2ml05} shows  a plot of $\tilde{Q}$ versus $\tilde{E}$ and that of $\tQ$ versus $\tilde{\omega}^{2}$  for equilibrium Q-ball solutions.
In the case of $\kappa=0$ there is one-to-one correspondence between $\tilde{Q}$ and $\tilde{E}$ while cusp 
structures appear in the case of $\kappa\ne0$, as shown in (a).
The maximum of $\tQ$ (labeled as  $A$ for $\kappa =0.006$) and the local minimum 
(labeled as $B$ for $\kappa =0.006$) appear in the case of gravitating Q-balls. 
At the point $B$, another cusp strcture appears and is far smaller than that at the point $A$. 
This sequences of cusp structure continue and we stopped calculation where the 4th cusp structure appears. 
Similar structures have been reported in \cite{KKL05,Grav-Q}. 
The $\tilde{Q}$-maximum for $\kappa =0.6$ is far smaller than that for $\kappa =0.006$.
Figure \ref{QEw2ml05}(b) shows that the stability criterion 
(\ref{Qw-condition}), which was established by perturbation on equilibrium Q-balls in flat spacetime,
cannot apply to gravitating Q-balls.

\begin{figure}[htbp]
\psfig{file=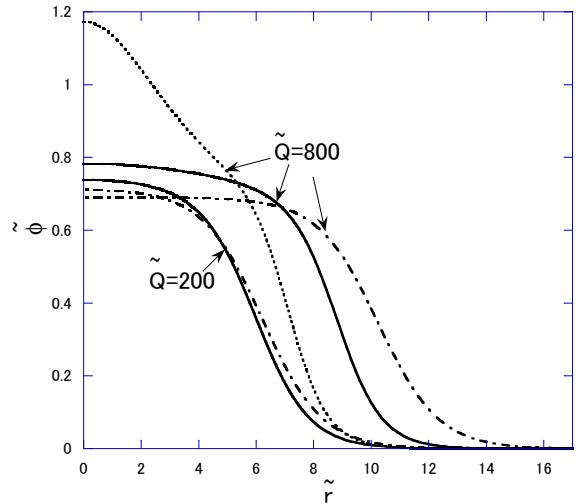,width=3in}  
\caption{$\tilde{\phi}$ as a function of $\tilde{r}$ for $\kappa =0$ (dot-dashed lines) and for $\kappa =0.006$
 (solid lines and a dotted line). 
For each $\kappa$, we take $\tilde{Q}=200$ and $800$.
\label{phicomp}}
\end{figure}

To see how gravity changes stability structure of equilibrium solutions, 
we show the profiles of the scalar field for $\kappa =0$ and those for $\kappa =0.006$ in Fig.~\ref{phicomp}. 
We find that gravitational effects reduce the Q-ball size and they become larger as $\tQ$ becomes larger. 
We also find that there are two solutions for $\tilde{Q}=800$ and $\kappa =0.006$. 
The solution with larger $\tilde{\phi}$ near the origin corresponds to the dotted line in 
Fig.~\ref{QEw2ml05}. We suppose that Q-balls with dotted lines 
cannot support itself and will collapse or disperse and the point $A$ would be the point 
where stability changes. Accordingly, there also exists the maximum of the Q-ball size due to 
gravity, as was pointed in \cite{multamaki}. 
We thus reasonably understand that the maximum of $\tilde{Q}$ for $\kappa =0.6$ is far smaller than that for $\kappa =0.006$.

\begin{figure}
\psfig{file=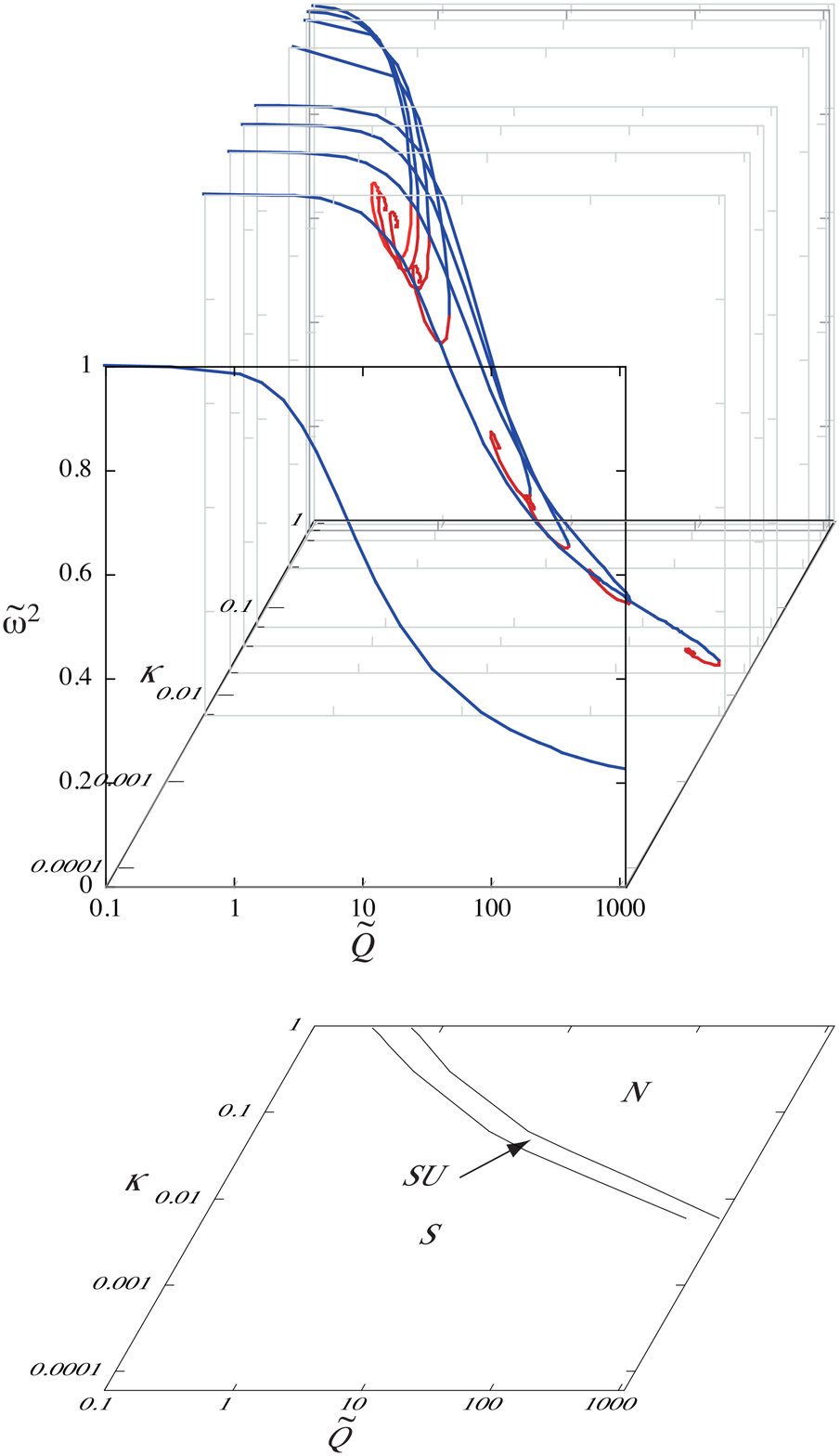,width=3in}  
\caption{\label{m6}
Structures of the {\it equilibrium spaces},
$M=\{(\to^2,\kappa,\tQ)\}$, and their catastrophe map, $\chi(M)$, 
into the {\it control planes}, $C=\{(\tilde{\mu}^2 ,\tQ)\}$, for $\tilde{\mu}^2 =\frac{5}{3}$.
Blue lines and red lines in $M$ 
represent stable and unstable solutions, respectively. 
In the regions denoted by S, SU and N on $C$,
there are one stable solution, one stable solution and one or more unstable solutions,
and no equilibrium solution, respectively, for fixed $(\kappa,\tQ)$.}
\end{figure}

Figure \ref{m6} shows the structures of the {\it equilibrium spaces}, 
${\cal M}=\{(\to^2,\kappa,\tQ)\}$, and their catastrophe map, $\chi({\cal M})$, 
into the {\it control planes}, $C=\{(\tilde{\mu}^2 ,\tQ)\}$, for $\tilde{\mu}^2 =\frac{5}{3}$. 
$\chi({\cal M})$ shows that in the regions denoted by S, SU and N on $C$,
there are one stable solution, one stable solution and one or more unstable solutions,
and no equilibrium solution, respectively, for fixed $(\kappa,\tQ)$. 
For example, for $\kappa =0.6$, which is the case chosen in Fig.~\ref{QEw2ml05}, 
we can confirm that unly a stable solution exists below $\tQ \sim 0.5$ while 
one stable solution and one or more unstable solutions exist in the region from 
$\tQ \sim 0.5$ to $\tQ \sim 0.9$. 
Here we demonstrate only the results for $\to>0$, that is, $\tQ>0$.
This does not mean $\to$ and $\tQ$ are always positive; 
the sign transformation $\to\ra-\to$ changes nothing but $\tQ\ra-\tQ$.
Stability of the solutions is determined by their energy $\tE$, as calculated in Fig.~\ref{QEw2ml05}. 

\begin{figure}[htbp]
\psfig{file=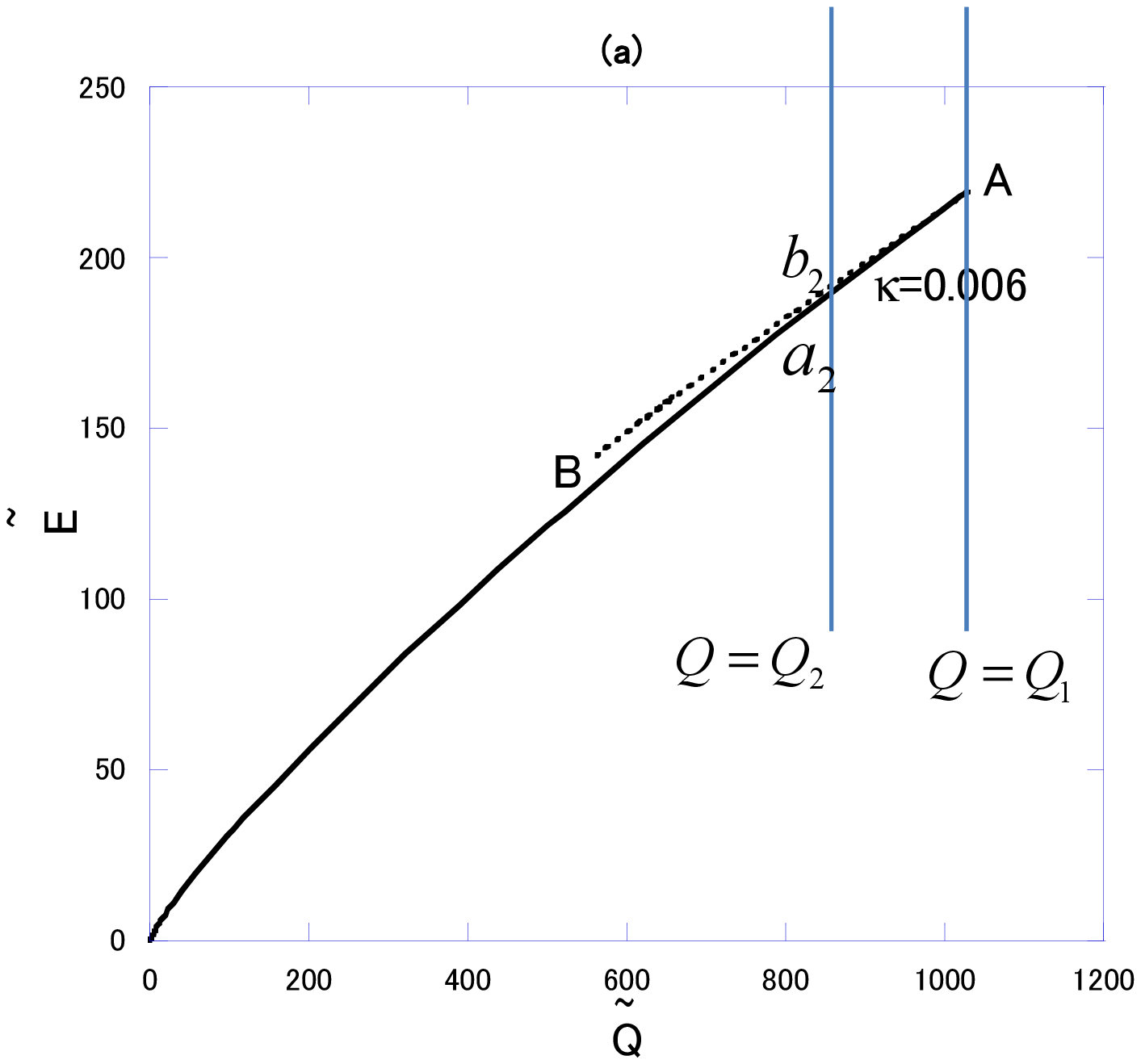,width=2in} \\
\psfig{file=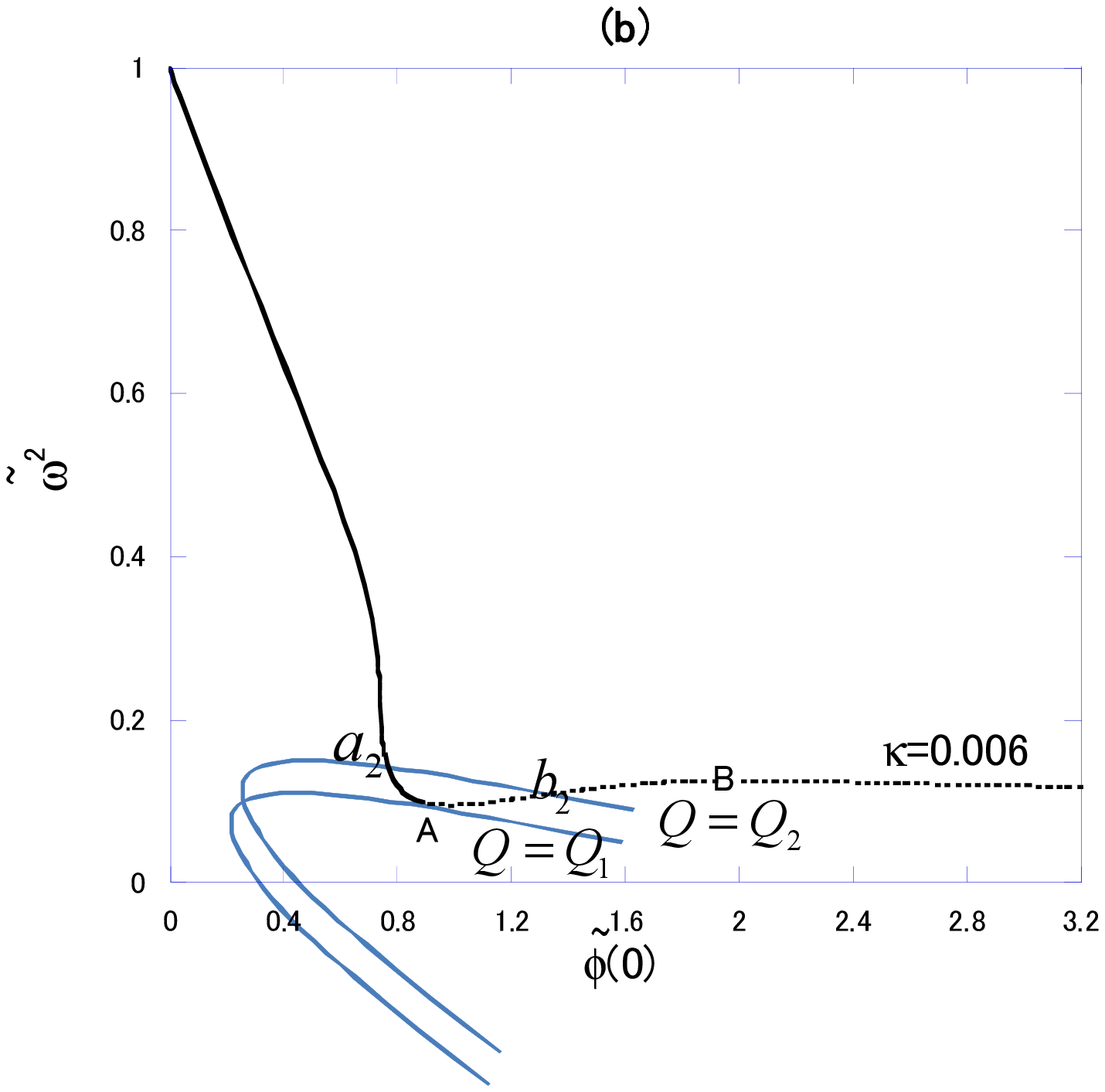,width=2in} \\
\psfig{file=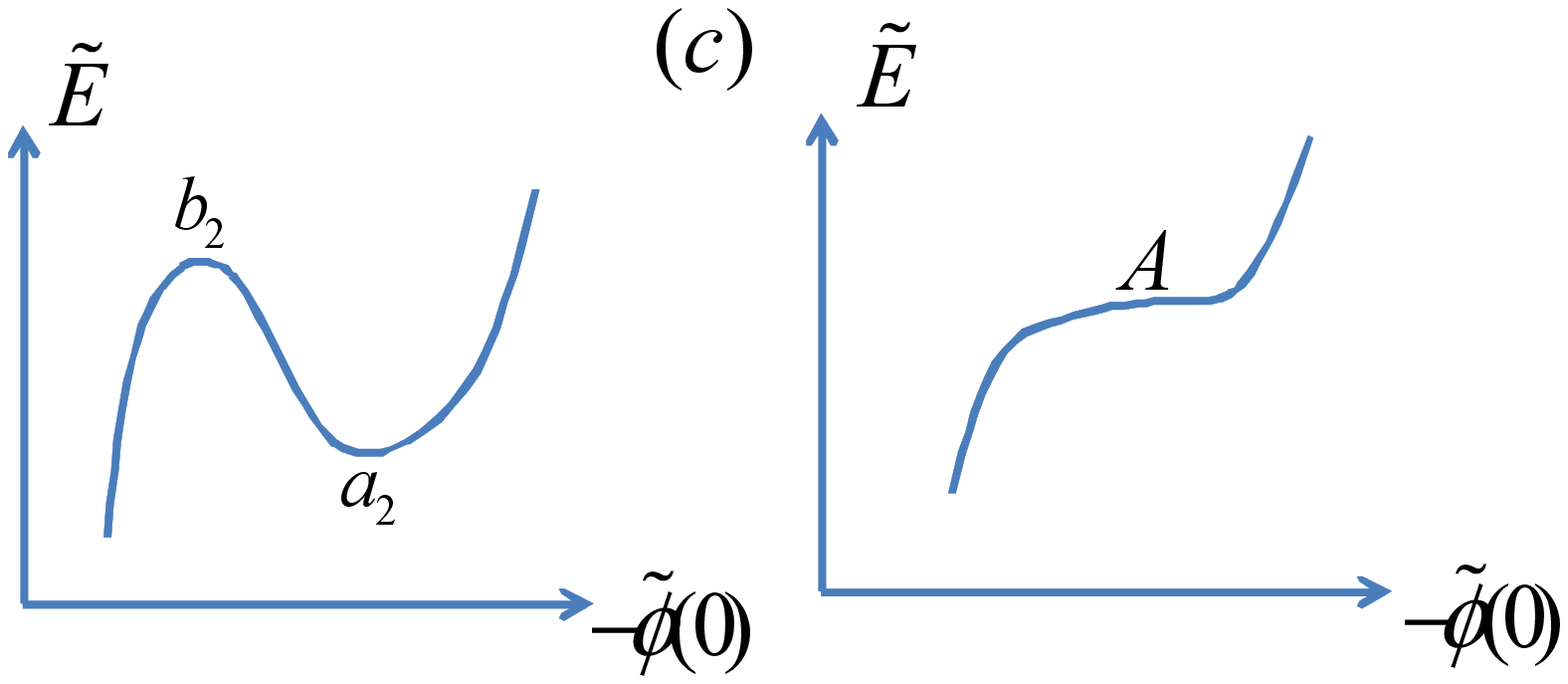,width=2in}
\caption{Stability interpretation via catastrophe theory for $\tilde{\mu}^2 =\frac{5}{3}$ 
for the gravitating case with $\kappa =0.006$ (The qualitative properties are same 
for other $\kappa$). The lines when we fix the control parameter $\tilde{Q}$ in (a) 
are depicted by the quadratic curves in (b). 
$a_2$ and $b_2$ are regarded as the potential minimum and the maximum in (c), respectively. 
\label{m06catas-k001} }
\end{figure}
\begin{figure}[htbp]
\psfig{file=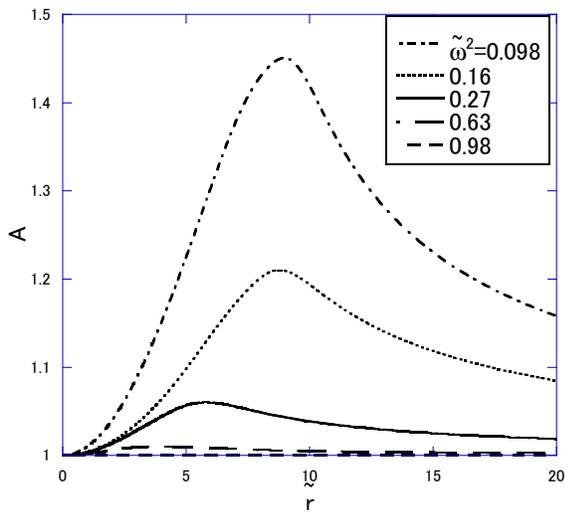,width=3in} 
\caption{Behavior of the metric $A$ for $\tilde{\mu}^2 =\frac{5}{3}$ 
for the gravitating case with $\kappa =0.006$. 
\label{k001A} }
\end{figure}
\begin{figure}[htbp]
\psfig{file=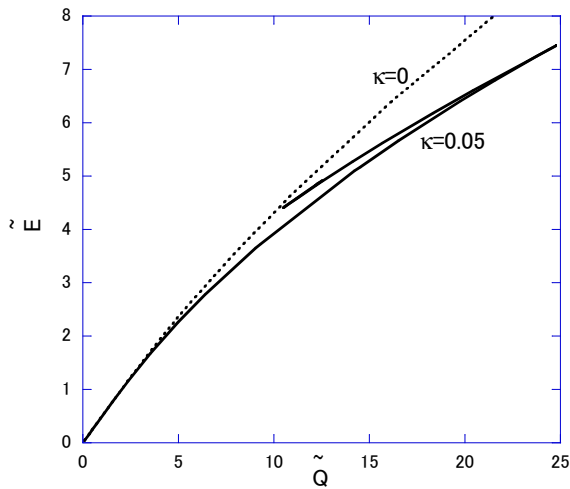,width=3in}  
\caption{$\tilde{Q}$-$\tilde{E}$ relation for 
$\tilde{\mu}^{2}=2$ in $V_{3}$ Model. We compare solutions for $\kappa=0$ and $0.05$. 
In the case of gravitating Q-balls ($\kappa\ne0$) cusp structures can be seen  
as in the case with $\tilde{\mu}^{2}=\frac{5}{3}$. 
\label{Fig6}}
\end{figure}

We explain our interpretation via catastrophe theory by exemplifying the case with 
$\tilde{\mu}^2 =\frac{5}{3}$ and $\kappa =0.006$ (Qualitative properties are not changed for other $\kappa$.). 
We identify $\tilde{Q}=$const.\ lines in Fig.~\ref{m06catas-k001} (a) with 
the quadratic curves in Fig.~\ref{m06catas-k001} (b). 
If we adopt the view point that stability changes at the point $A$ 
as we mentioned above and observe Fig.~\ref{m06catas-k001} (b), 
we notice that $(-\tilde{\phi}(0))$ is more appropriate for a {\it behavior variable} 
than $\tilde{\omega}^{2}$. Then, as we show in (c), $a_{2}$, $b_2$ and $A$ can be interpreted 
as the potential minimum, maximum and the inflection point, respectively. 

This case can be understood using the fold catastrophe 
$f(u)=u^{3}+tu$ where $u$ and $t$ are the {\it bahavior variable} identified with  
$(-\tilde{\phi}(0))$ and 
the {\it control parameter} identified with $\tilde{Q}$, respectively. 

We should reveal what causes the difference from the flat case. 
We show behavior of the metric $A$ for the solutions with $\tilde{\omega}^{2}=0.098$, 
$0.16$, $0.27$, $0.63$ and $0.98$ for $\kappa =0.006$ in Fig.~\ref{k001A}. We should notice its relation to 
Fig.~\ref{m06catas-k001} (b). Naively speaking, solutions having larger $|A-1|$ 
at its peak have larger $(-\tilde{\phi}(0))$. We have confirmed that, independent of $\kappa$, 
solutions having $|A-1|\sim 1$ correspond to solutions expressed by dotted lines in 
Fig.~\ref{QEw2ml05}. Therefore, we can suppose that the intrinsic difference from 
the flat case can be characterized by $|A-1|$. 

We have also confirmed that these properties can be seen for other $\tilde{\mu}^2 $. 
As an example, we exhibit $\tilde{Q}$-$\tilde{E}$ relation for $\tilde{\mu}^{2}=2$ in Fig.~6. 
In the case of gravitating Q-balls ($\kappa\ne0$) cusp structures can be seen  
as in the case with $\tilde{\mu}^{2}=\frac{5}{3}$. 
Other diagrams and related properties also resemble to the case with $\tilde{\mu}^{2}=\frac{5}{3}$. 

\begin{figure}[htbp]
\psfig{file=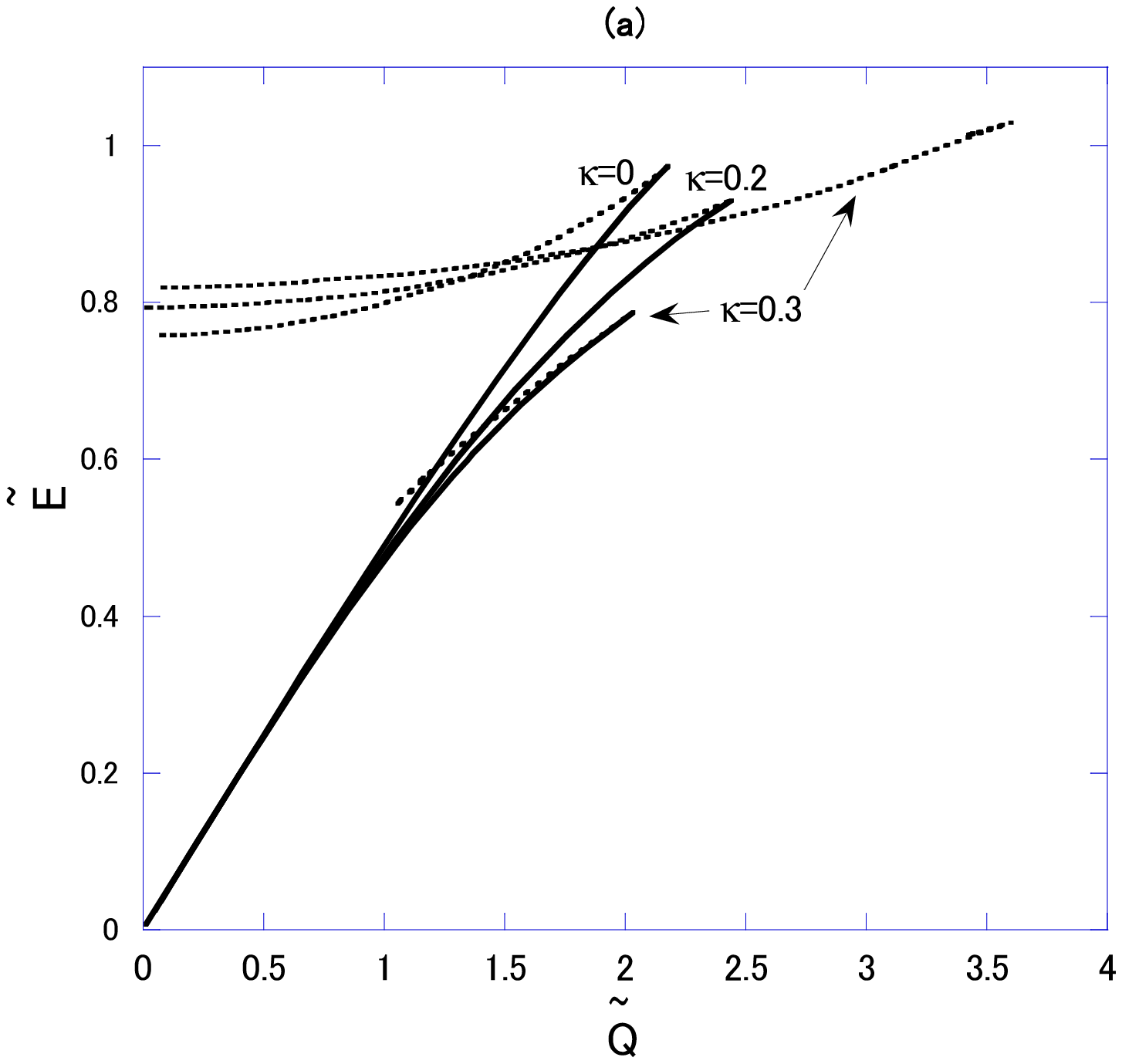,width=3in}  \\
\psfig{file=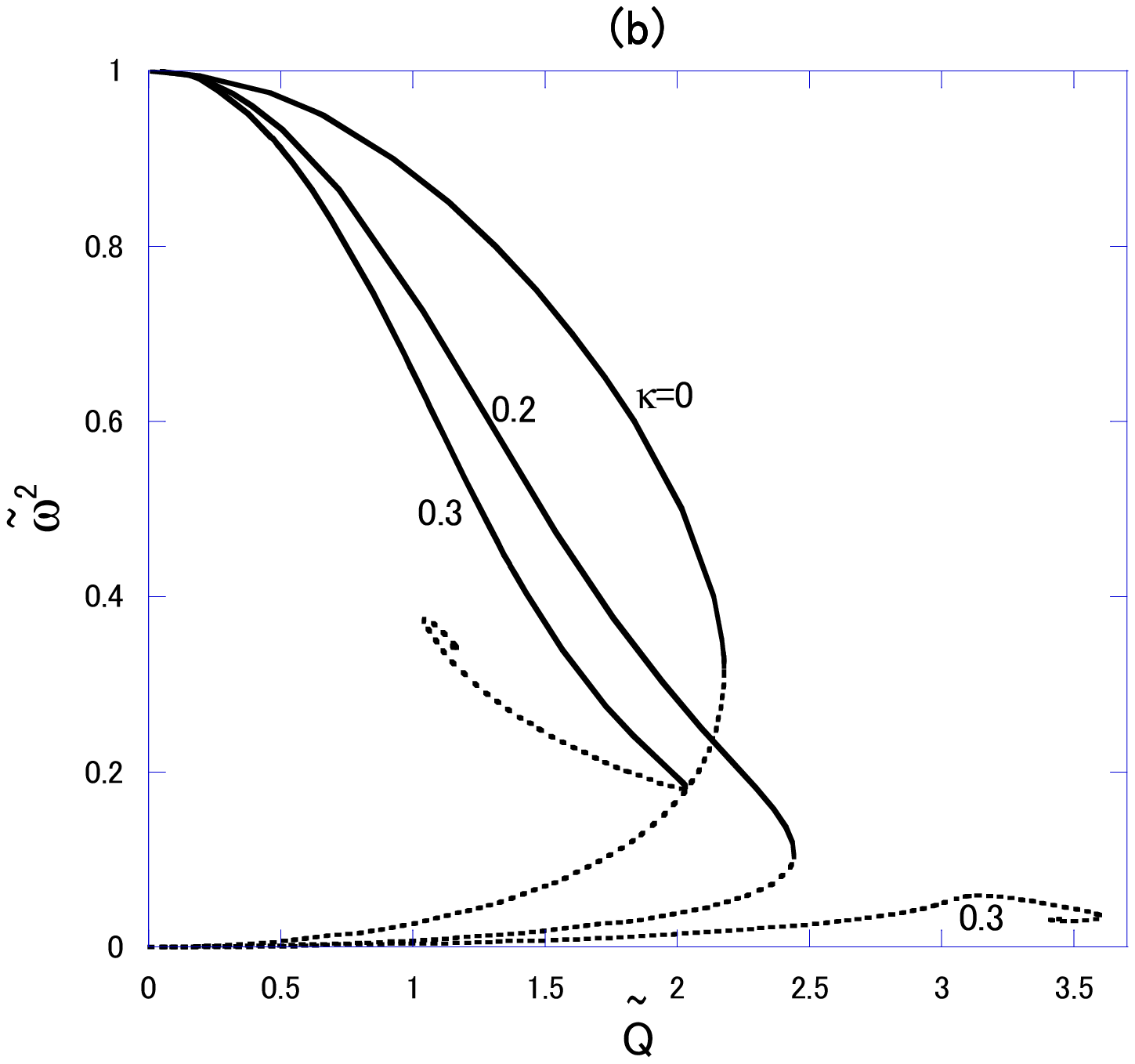,width=3in}
\caption{(a) $\tilde{Q}$-$\tilde{E}$ relation and (b) $\tilde{Q}$-$\tilde{\omega}^{2}$ relation for 
$\tilde{\mu}^2 =5$. Intrinsic difference can be seen for $\kappa =0.3$. 
\label{QEw2m02} }
\end{figure}

\subsection{Gravitating Q-balls for $\tilde{\mu}^2>2$}

\begin{figure}
\psfig{file=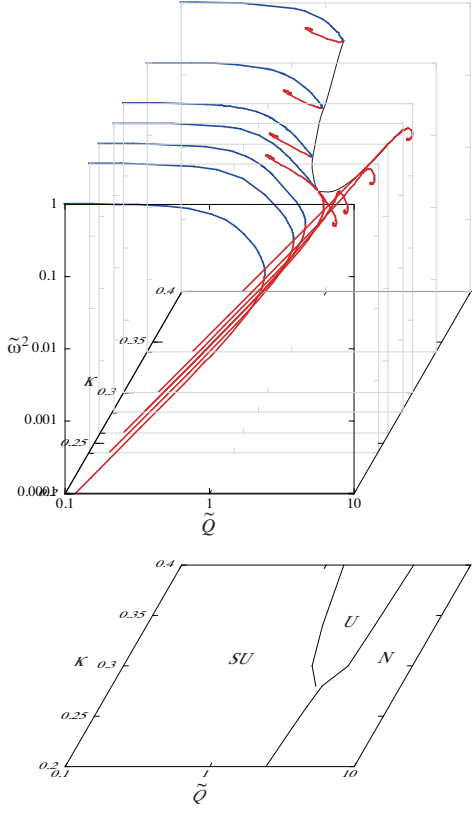,width=3in}  
\caption{\label{m2}
Structures of the {\it equilibrium spaces},
{\cal M}$=\{(\to^2,\kappa,\tQ)\}$, and their catastrophe map, $\chi(${\cal M}$)$, 
into the {\it control planes}, $C=\{(\tilde{\mu}^2 ,\tQ)\}$, for $\tilde{\mu}^2 =5$.
Blue lines and red lines in M 
represent stable and unstable solutions, respectively. 
In the regions denoted by S, SU and N on $C$,
there are one stable solution, one stable solution and one or more unstable solutions,
and no equilibrium solution, respectively, for fixed $(\kappa,\tQ)$.}
\end{figure}

Next, we fix $\tilde{\mu}^2 =5$ as an example of $\tilde{\mu}^2>2$.
Figure \ref{QEw2m02} shows a plot of $\tilde{Q}$ versus $\tilde{E}$ and that of $\tilde{Q}$ versus 
$\tilde{\omega}^{2}$ for $\kappa=0$, $0.2$ and $0.3$.
In flat spacetime ($\kappa=0$), contrary to the case of $\tilde{\mu}^2 < 2$, 
$\tilde{Q}$ has a maximum $\tQ_{{\rm max}}$, and for each $\tQ(<\tQ_{{\rm max}})$ there is 
one stable solution and one unstable solution~\cite{SakaiSasaki}. 
The solution of the $\tilde{Q}$-maximum coincides with that of the $\tilde{E}$-maximum. 
Figure  \ref{QEw2m02} indicates that gravitational effects for the case of $\kappa=0.2$ is not so large, 
compared with the case of $\kappa=0.006$ and $\tilde{\mu}^2 =\frac{5}{3}$ in Fig.\ \ref{QEw2ml05}.
This is simply due to the smallness of the gravitational mass $2\tE$.

In the case of $\kappa =0.3$, we find a cusp structure similar to that 
for $\tilde{\mu}^2 =\frac{5}{3}$. We call it a degenerate cusp structure. 
We also confirmed that solutions written by a dotted line for $\kappa =0.3$ 
have the metric $|A-1|\sim 1$ at its peak. Then, again, we suppose that solutions with 
strong gravity $|A-1|\sim 1$ show similar cusp structure regardless of the potential parameters. 

\begin{figure}[htbp]
\psfig{file=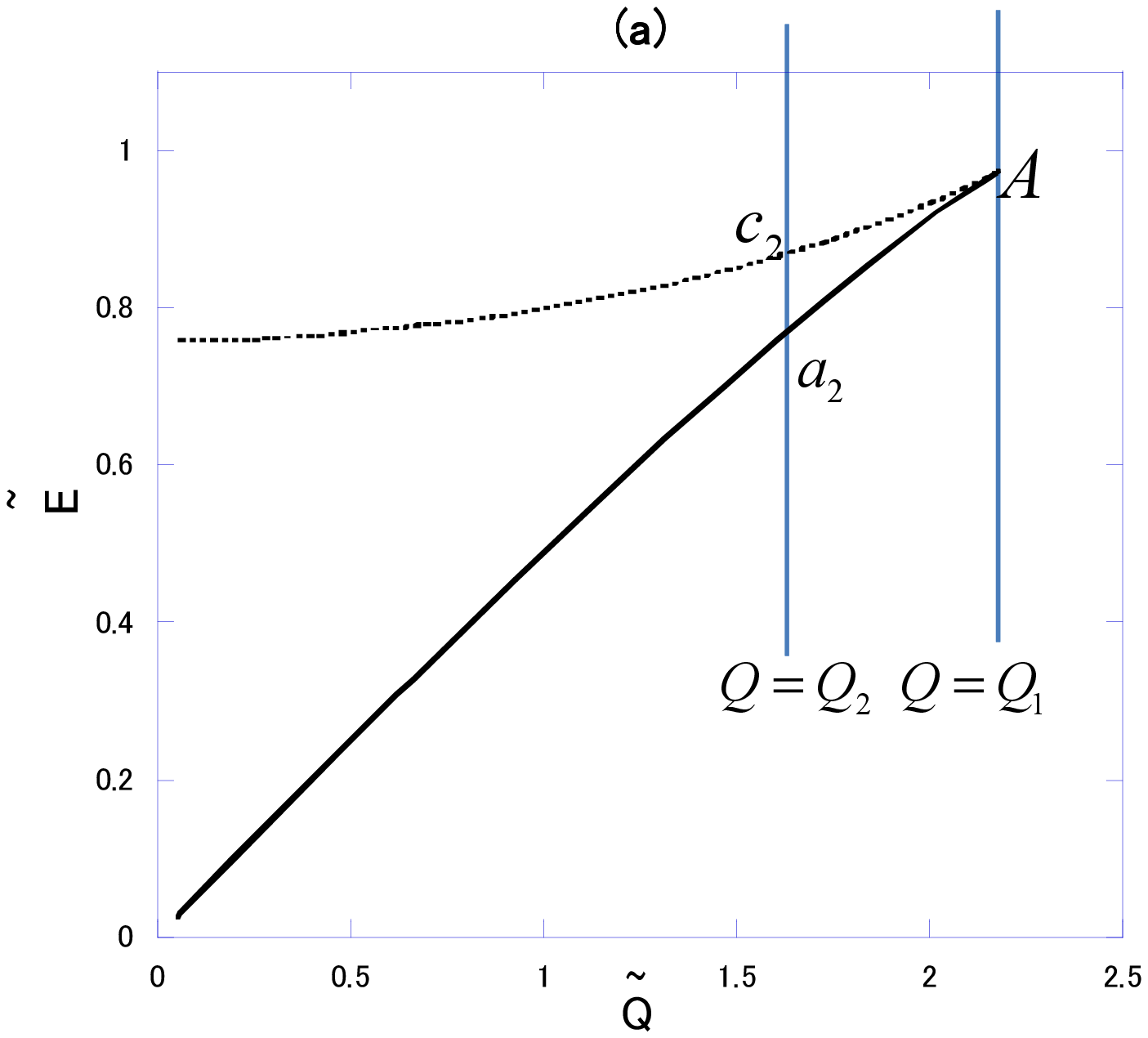,width=3in} \\
\psfig{file=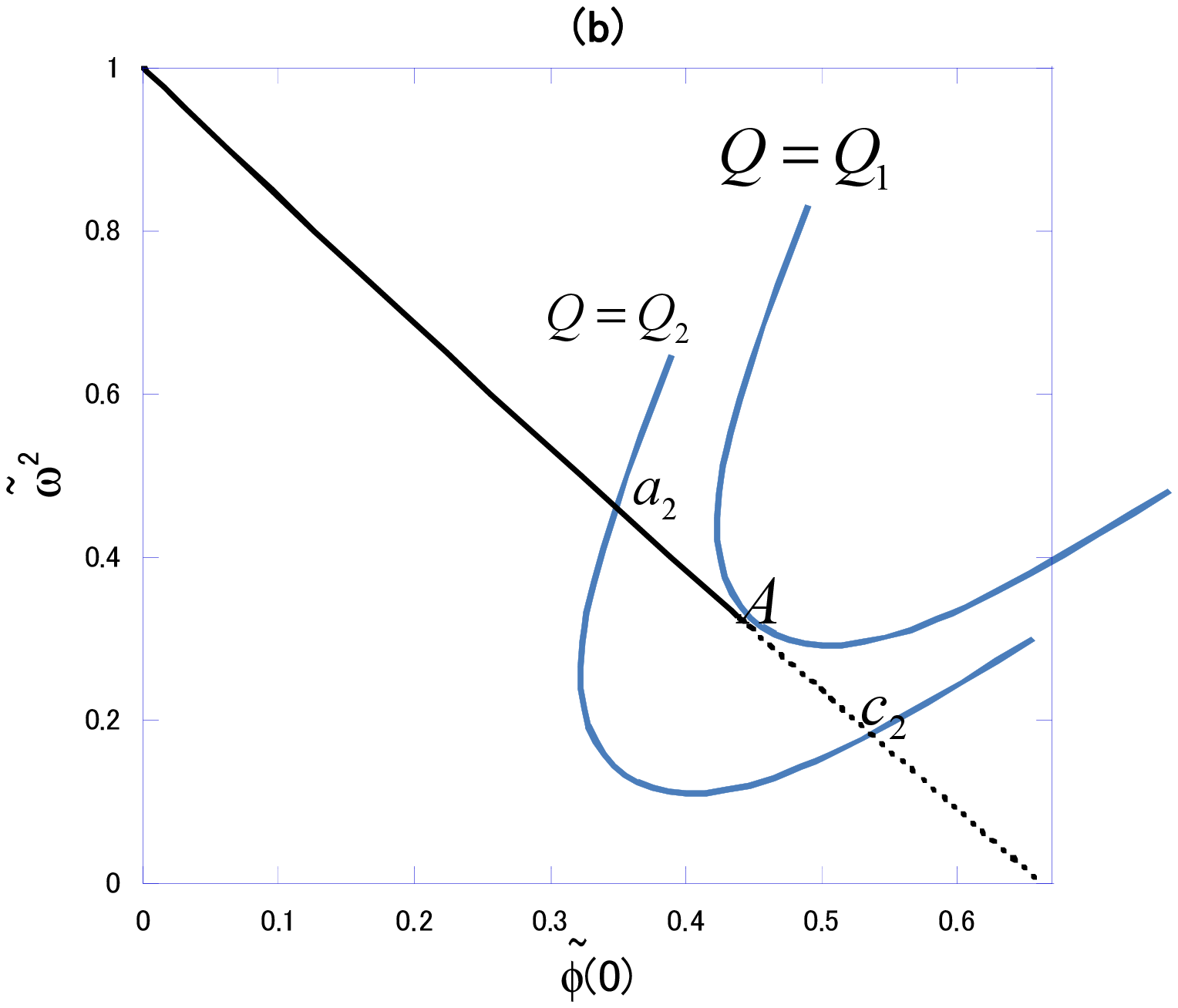,width=3in} \\
\psfig{file=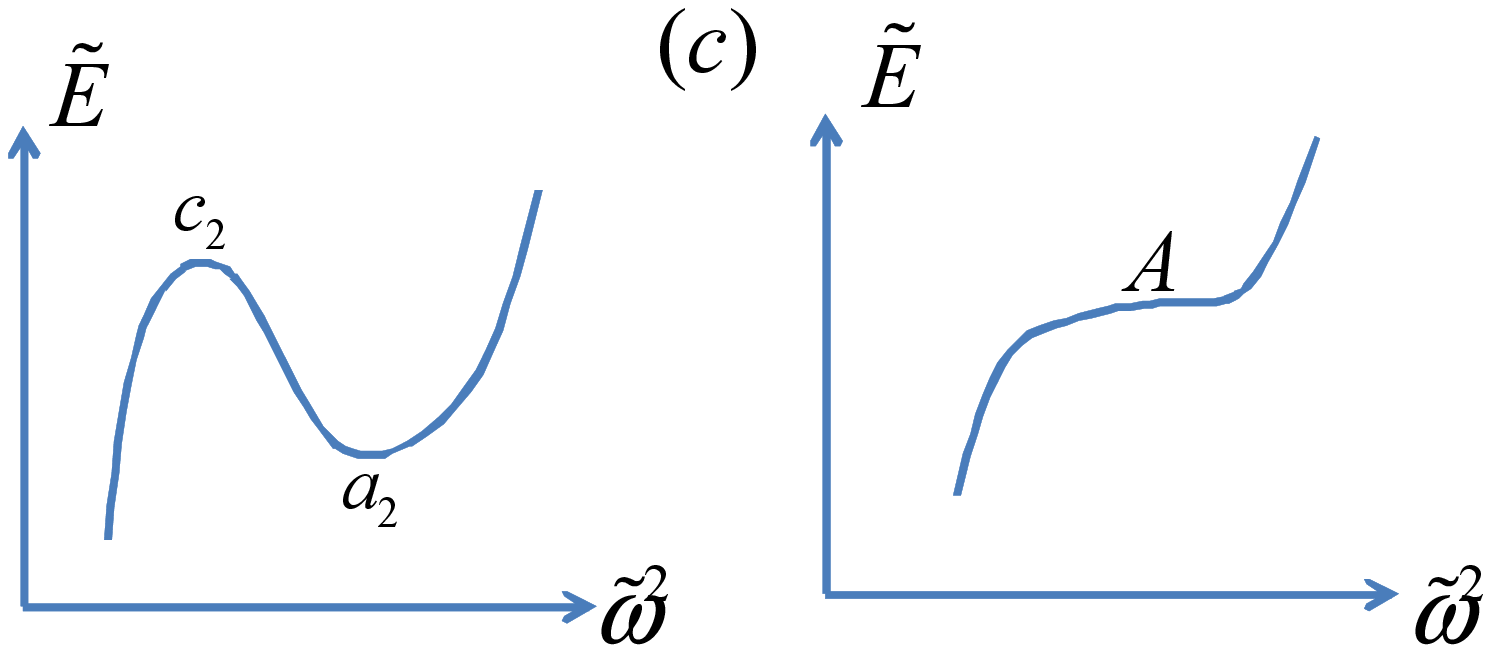,width=3in}
\caption{Stability interpretation via catastrophe theory for $\tilde{\mu}^2 =5$ 
for the flat case. The lines when we fix the control parameter $\tilde{Q}$ in (a) 
are depicted by the quadratic curves in (b). The points $a_2$ ($c_2$) are supposed 
to be stable (unstable) since it is expressed as the potential minimum (maximum) shown in (c). 
\label{m02catas-flat} }
\end{figure}

In addition, we have solutions with another sequences of cusp structures at far 
larger  $\tilde{Q}$ and $\tilde{E}$ around $\tQ \sim 3.5$. 
We call the former sequence the low energy branch, while the latter sequence the high energy branch.
The low energy  branch is similar to the sequence of $\tilde{\mu}^2 =\frac{5}{3}$ in Fig.\ \ref{QEw2ml05},
which suggests that the stability structure is also similar.
On the other hand, we suppose that any solution in the high energy branch is unstable since it 
basically corresponds to that written by a dotted line for $\kappa =0$, which is already shown as unstable. 
In the regions denoted by S, SU and N on $C$,
there are one stable solution, one stable solution and one or more unstable solutions,
and no equilibrium solution, respectively, for fixed $(\kappa,\tQ)$. 
Figure \ref{m2} shows the structures of the {\it equilibrium spaces}, 
${\cal M}=\{(\to^2,\kappa,\tQ)\}$, and their catastrophe map, $\chi({\cal M})$, 
into the {\it control planes}, $C=\{(\tilde{\mu}^2 ,\tQ)\}$, for $\tilde{\mu}^2 =5$.

We see that one sequence diverges into two branches at $\kappa\approx 0.28$,
above which each branch exposes spiral structures. 
The upper branch in Fig.\ \ref{m2} is analogous to the solution sequence for $\tilde{\mu}^2 =\frac{5}{3}$ in Fig.\ \ref{m6}.
This also indicates that, as the strength of gravity becomes large, the dependence on the potential shape diminishes.

\begin{figure}[htbp]
\psfig{file=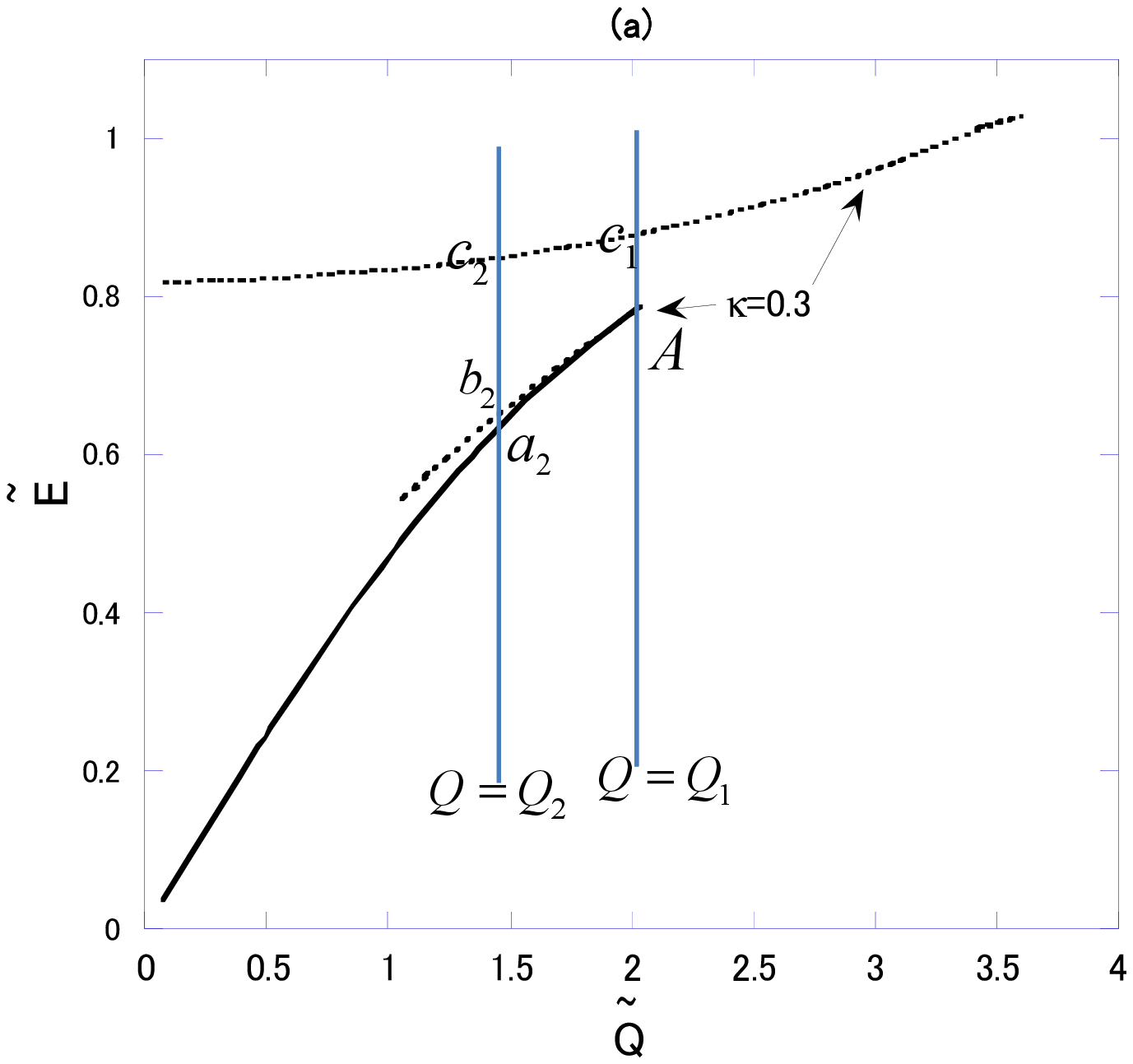,width=3in} \\
\psfig{file=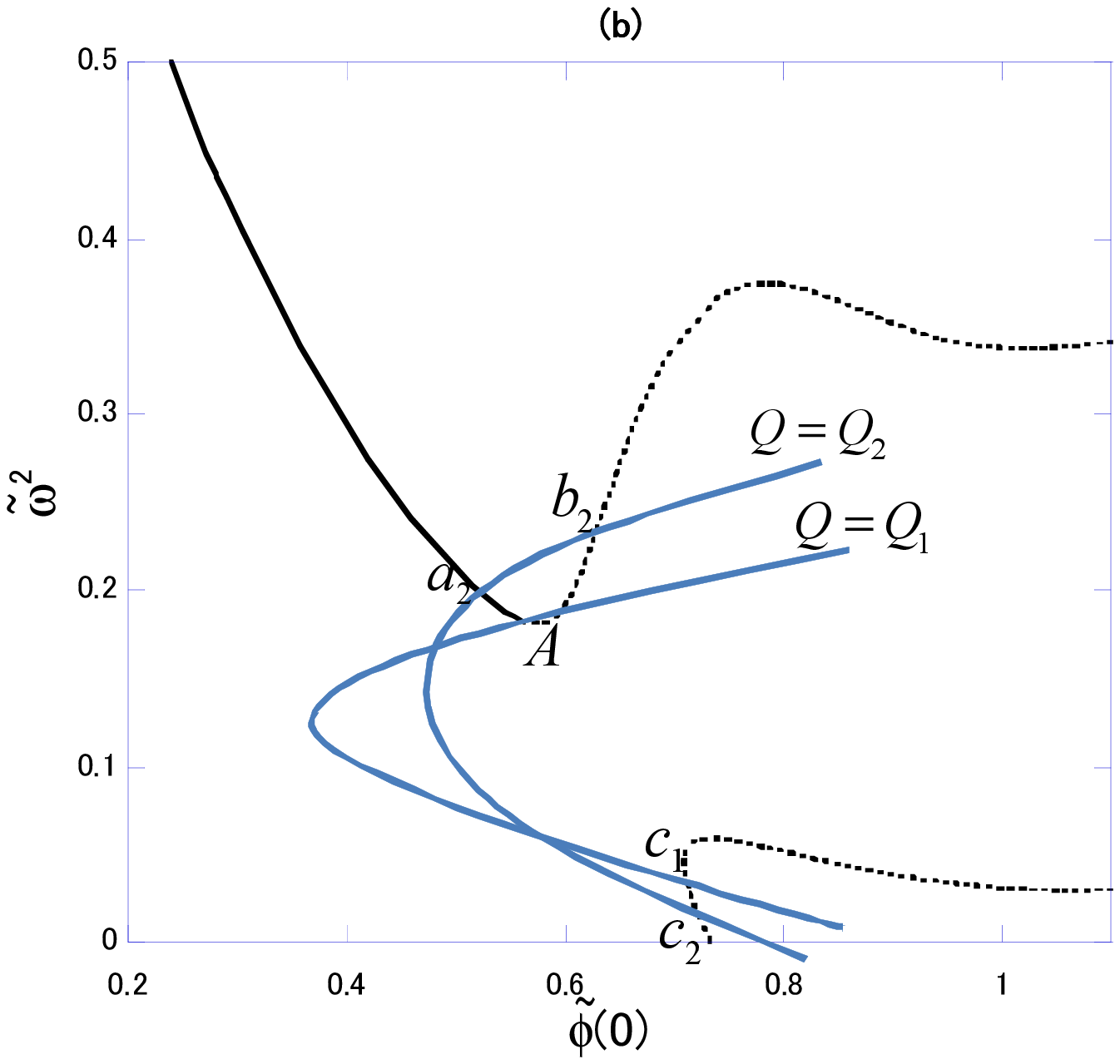,width=3in} 
\caption{Stability interpretation via catastrophe theory for $\tilde{\mu}^2 =5$ 
for the gravitating case with $\kappa =0.3$ (The qualitative properties are same 
for other $\kappa > 0.28$. The case for $\kappa < 0.28$ can be understood as 
in the same way in the flat case.). 
The lines when we fix the control parameter $\tilde{Q}$ in (a) 
are depicted by the quadratic curves in (b). 
In this case, potential plane is interpreted as Fig.~\ref{m06catas-k001} (c). 
\label{m02catas-k15} }
\end{figure}
\begin{figure}[htbp]
\psfig{file=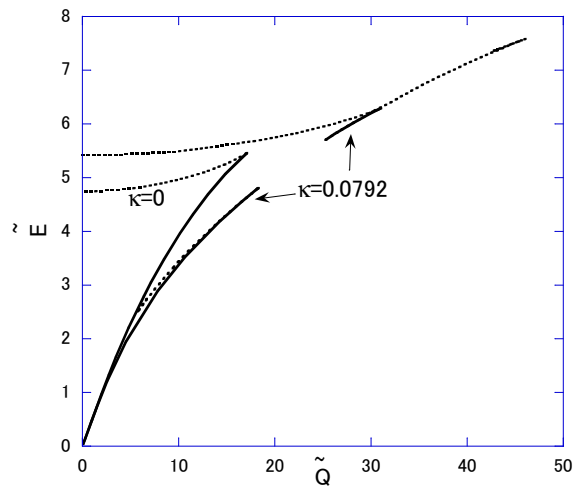,width=3in}  
\caption{$\tilde{Q}$-$\tilde{E}$ relation for 
$\tilde{\mu}^2 =\frac{5}{2}$ which is qualitatively same as the case with $\tilde{\mu}^2 =5$. 
\label{Fig11} }
\end{figure}

We can understand its catastrophe type using fold catastrophe as the case with $\tilde{\mu}^2 =\frac{5}{3}$. 
We also identify $\tilde{Q}=$const.\ laines in Fig.~\ref{m02catas-flat} (a) with 
the quadratic curves in Fig.~\ref{m02catas-flat} (b) where flat solutions are shown. 
In this case, there is a one-to-one correspondence between $\tilde{\omega}^{2}$ and $\tilde{\phi}(0)$.
Therefore, both variables can be behavior variables as shown in Fig.~\ref{m02catas-flat} (c). 

We also exhibit the corresponding figures with $\kappa =0.3$ 
in Fig.~\ref{m02catas-k15}. $\tilde{\omega}^{2}$-$(-\tilde{\phi}(0))$ 
relation changes drastically 
as shown in Fig.~\ref{m02catas-k15} (b). This figure suggests that we should use 
$(-\tilde{\phi}(0))$ as a behavior variable similar to Fig.~\ref{m06catas-k001} (c) near the point $A$. 
As for the high energy branch, since the point $c_{2}$ corresponds to that in 
Fig.~\ref{m02catas-flat} (c), we can naturally suppose that this branch is unstable. 

For completeness, in Fig.~\ref{Fig11}, we also show $\tilde{Q}$-$\tilde{E}$ relation for 
$\tilde{\mu}^2 =\frac{5}{2}$ with $\kappa =0$ and $0.0792$ which is qualitatively similar to 
the case with $\tilde{\mu}^2 =5$. The solid and the dotted lines correspond to the stable and 
the unstable solutions, respectively. As for the case $\tilde{\mu}^2 =\frac{5}{2}$ 
with $\kappa =0.0792$ the only difference from the case $\tilde{\mu}^2 =5$ with 
$\kappa =0.3$ is that there are stable solutions in the high energy branch. 
However, this is not the intrinsic difference. Actually, for the case $\tilde{\mu}^2 =5$ with 
$\kappa \sim 0.28$, there are stable solutions in the high energy branch which 
correspond to the point $a_{2}$ as in the low energy branch. 
As we can expect from this diagram, other relations also resemble to those with $\tilde{\mu}^2 =5$. 

\subsection{Boson stars}

\begin{figure}[htbp]
\psfig{file=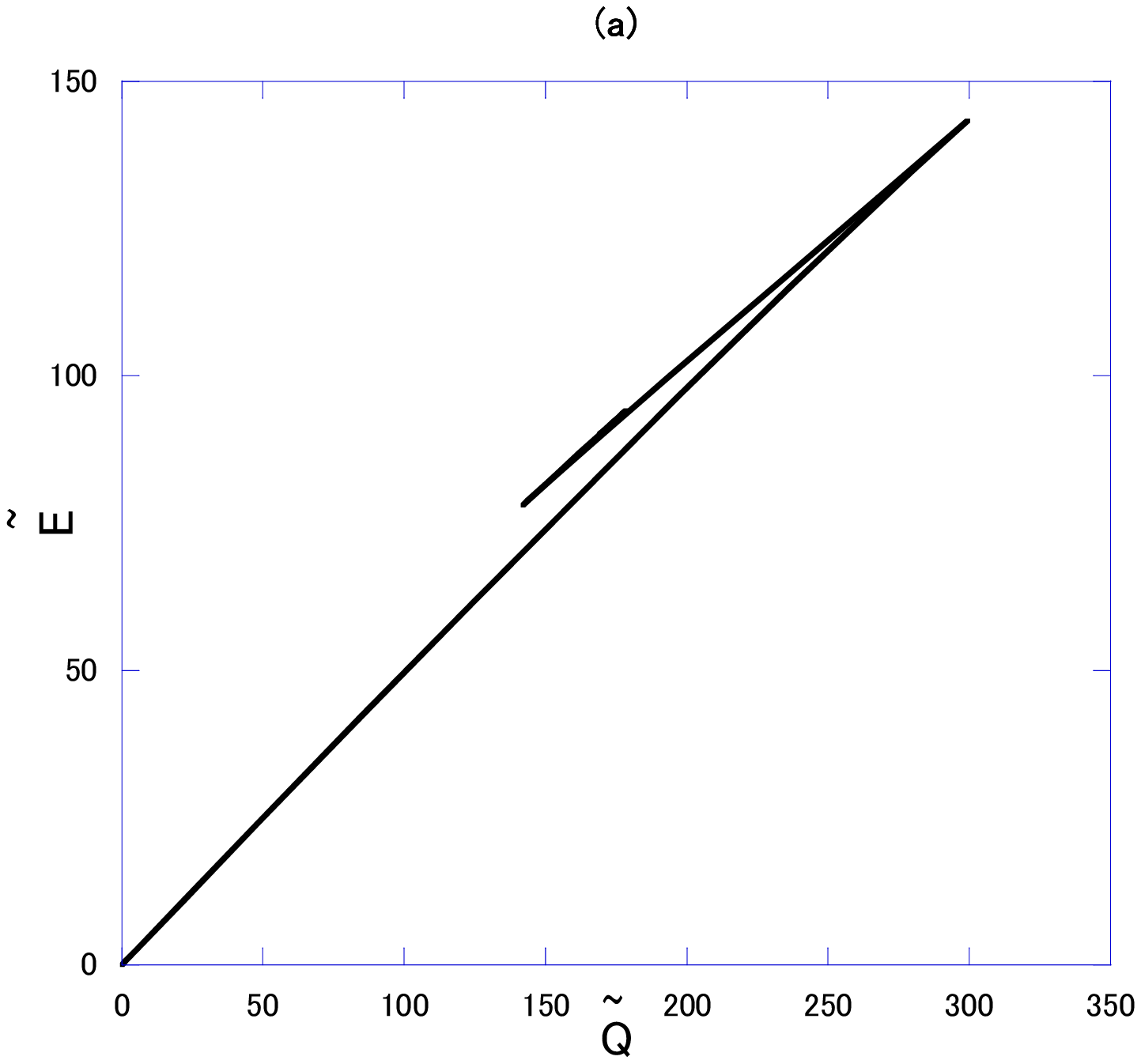,width=3in} \\
\psfig{file=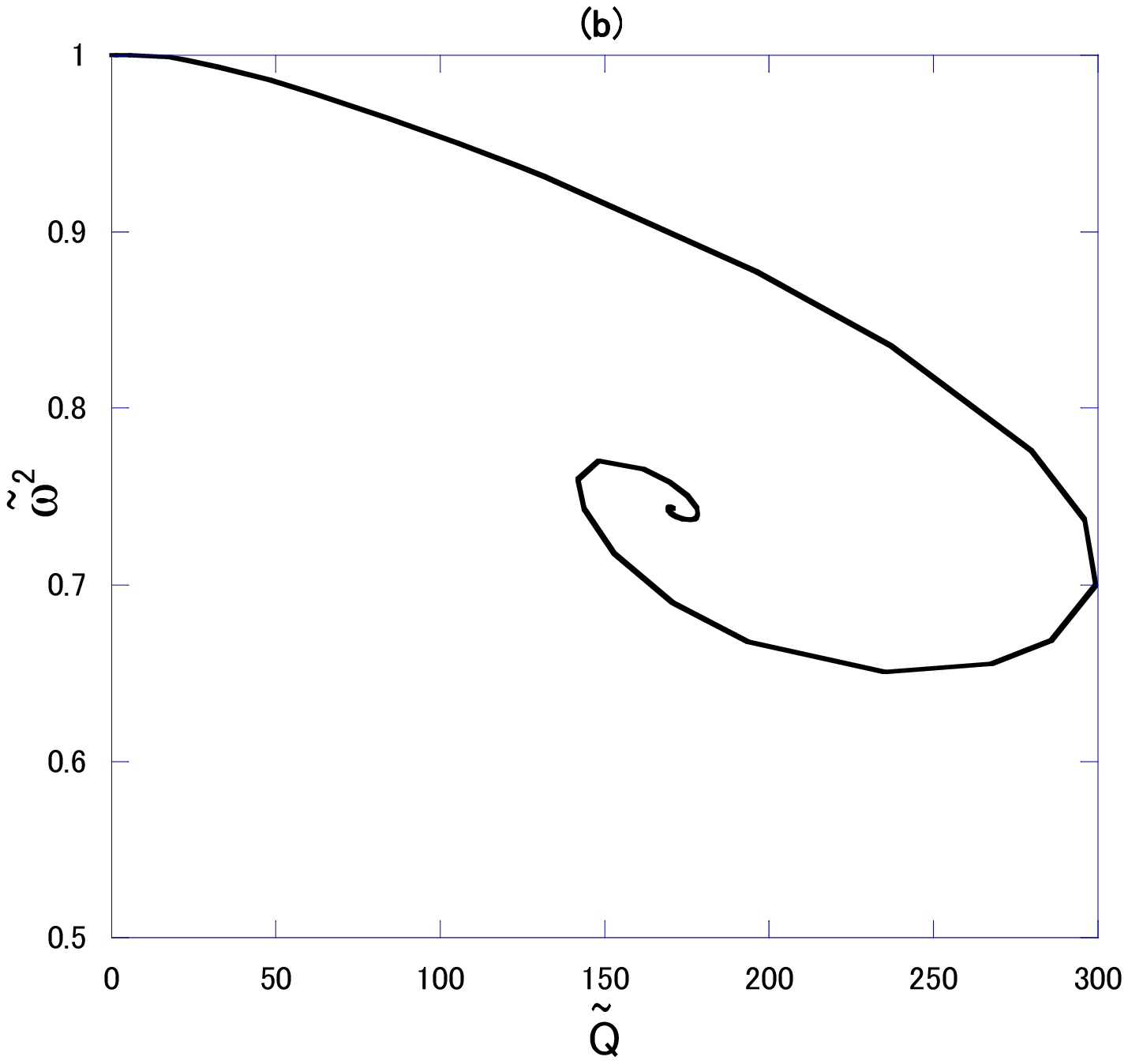,width=3in}
\caption{(a)$\tilde{Q}$-$\tilde{E}$, (b)$\tilde{Q}$-$\tilde{\omega}^{2}$ relations
in the model of boson stars for $\kappa =0.006$.
\label{QEBS}}
\end{figure}

Now we discuss boson stars with the potential (\ref{V3}) with $\mu =0$.
Figure \ref{QEBS} shows plots of (a) $\tQ$-$\tE$, (b) $\tQ$-$\to^2$ for equilibrium solutions of boson stars. 
Degenerate cusp and spiral structures are seen as in the case of gravitating Q-balls 
for $\tilde{\mu}^2 =\frac{5}{3}$ and those for $\tilde{\mu}^2 =5$ with $\kappa > 0.28$. 

\begin{figure}
\psfig{file=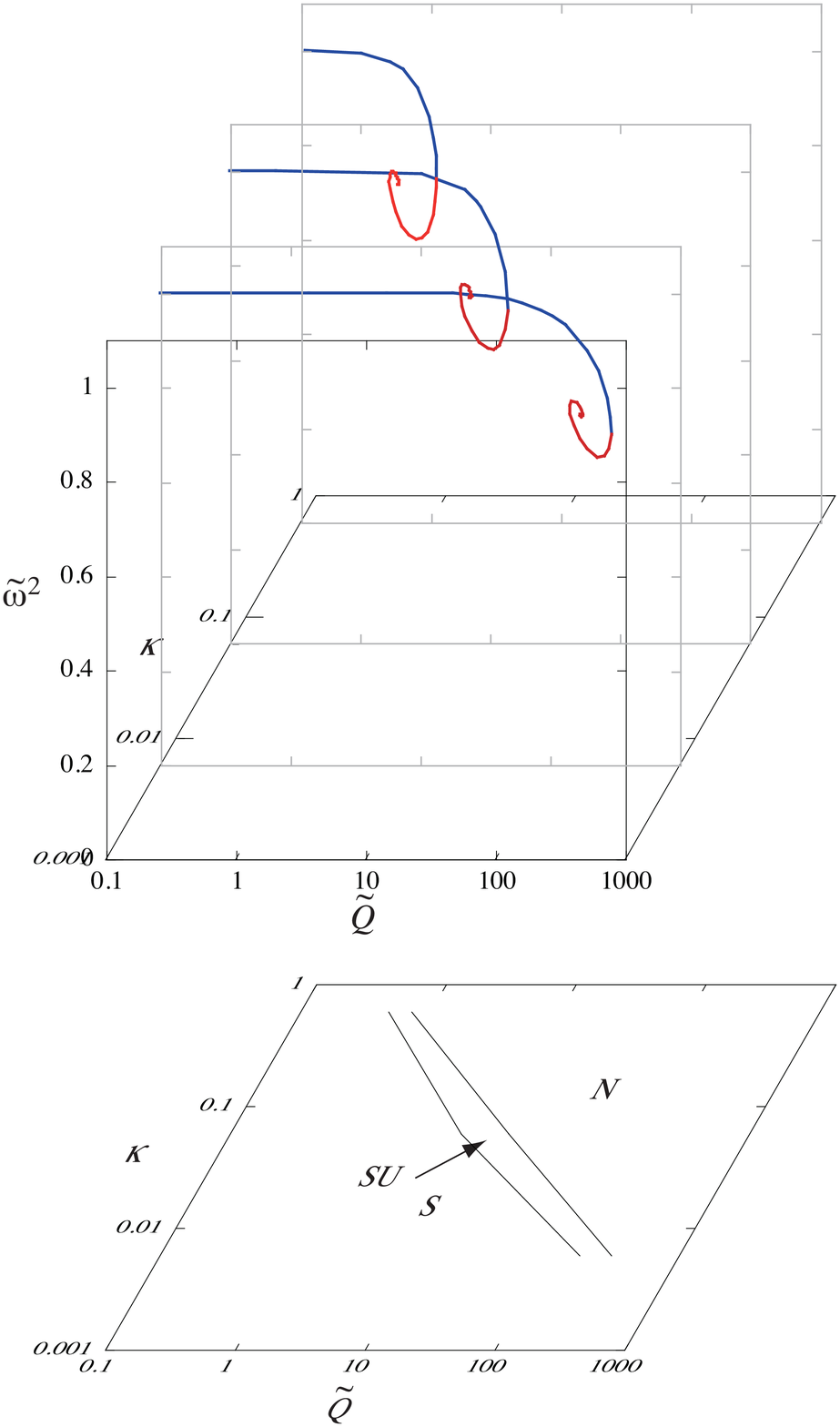,width=3in}  
\caption{\label{BS}
Structures of the {\it equilibrium spaces},
$M=\{(\to^2,\kappa,\tQ)\}$, and their catastrophe map, $\chi(M)$, 
into the {\it control planes}, $C=\{(\tilde{\mu}^2 ,\tQ)\}$, for $\tilde{\mu}^2 =0$. 
In the regions denoted by S, SU and N on $C$,
there are one stable solution, one stable solution and one or more unstable solutions,
and no equilibrium solution, respectively, for fixed $(\kappa,\tQ)$.}
\end{figure}

Figure \ref{BS} shows the structures of the {\it equilibrium spaces} and their catastrophe map
$\chi({\cal M})$ into the {\it control planes} for $\tilde{\mu}^2 =0$. 
In the regions denoted by S, SU and N on $C$,
there are one stable solution, one stable solution and one or more unstable solutions,
and no equilibrium solution, respectively, for fixed $(\kappa,\tQ)$.
We see that qualitative characteristics of the equilibrium space and its catastrophe type are the same 
as those for $\tmu^2 =5/3$.
We have confirmed $|A-1|\sim 1$ at its peak in the solutions corresponding to the spiral curves or near the stability change points.
We therefore conclude that, if gravity is so strong as $|A-1|\sim 1$ at its peak, 
catastrophic structures of Q-balls approach those of boson stars, regardless of the potential shape.

\section{Conclusion and discussion}
We have reanalyzed stability of gravitating Q-balls for a $V_{3}$ model and boson stars 
for a $V_{3}$ model with $\mu =0$. 
For solutions with $|g^{rr}-1|\sim 1$ at its peak, stability of Q-balls has 
been lost regardless of the potential parameters. As a result, phase relations, such as 
$\tQ$-$\tilde{E}$, approach those of boson stars, 
which tell us an unified picture of Q-balls and boson stars. 

Therefore, if we discuss the possibility 
of Q-balls or boson stars as dark matter candidates, our work 
would be useful. This work should 
also be extended to the $V_{4}$ model which will appear in our companion paper. 

\acknowledgements
We would like to thank Kei-ichi Maeda for useful discussion and for 
continuous encouragement. The numerical calculations were carried out 
on SX8 at  YITP in Kyoto University.
This work is in part supported by MEXT Grant-in-Aid for Scientific Research (C) No.\ 18540248.



\begin{thebibliography}{99}
\bibitem{footnote}
In our previous paper~\cite{SakaiSasaki}, we rescaled the quantities as 
\bea
\tt_b\equiv{\mu\over\sqrt{\lambda}}t,~~
\tr_b\equiv{\mu\over\sqrt{\lambda}}r,
&&\to_b\equiv{\sqrt{\lambda}\over\mu}\omega,~~
\tm_b\equiv{\sqrt{\lambda}\over\mu}m,\nn
\tp_b\equiv{\lambda\over\mu}\phi,~~
&&\tv_b\equiv{\lambda^3\over\mu^4}V_{3}={1\over2}\tp_b^2-\tmu\tp_b^3+\tp_b^4,\nn
\kappa_b\equiv{\mu^2G\over\lambda^2},~~
&&\tQ_b\equiv \lambda Q,~~\tE_b\equiv{\lambda^{{3\over2}}\over\mu}E\ .
\eea
For this reason, present quantities are written using the previous quantities as 
\bea
&&\tilde{\mu}={1\over\tm_b},~~
\to={\to_b\over\tm_b},~~
\kappa=\tm_b^2\kappa_b,\nn
&&\tp={\tp_b\over\tm_b},~~
\tQ=\tQ_b. 
\eea
As for the total energy $E$, care should be taken. In our previous paper, 
only $H_{\phi}$ has been considered while $H_{\phi}+H_{\rm G}$ 
are considered, in the present paper, to define the total energy for including 
the gravitating case. Because of this difference, in flat case, 
the total energy in the present paper is half of that in the previous paper. 
Therefore, we obtain the relation as 
\bea
\tE={\tE_b\over 2\tm_b}.~~
\eea
\end{thebibliography}
\end{document}